\documentclass[10pt,letterpaper,twocolumn]{article} 

\usepackage{ol2}
\usepackage[draft]{hyperref}
\usepackage{amsmath}
\usepackage{amsmath}
\usepackage{amssymb,graphicx}
\usepackage{cite}
\usepackage{bm}

\begin{document}

\newcommand{\vett}[1]{\mathbf{#1}}
\newcommand{\uvett}[1]{\hat{\vett{#1}}}
\newcommand{\beq}{\begin{equation}}
\newcommand{\eeq}{\end{equation}}
\newcommand{\barr}{\begin{eqnarray}}
\newcommand{\earr}{\end{eqnarray}}

\twocolumn[ 

\title{Generalized Bessel beams with two indices}

\author{Marco Ornigotti$^{1*}$ and Andrea Aiello$^{2,3}$}
\address{$^{1}$Institute of Applied Physics, Friedrich-Schiller University, Jena, Max-Wien Platz 1, 07743 Jena, Germany\\
$^{2}$Max Planck Institute for the Science of Light, G$\ddot{u}$nther-Scharowsky-Strasse 1/Bau24, 91058 Erlangen, Germany\\
$^{3}$Institute for Optics, Information and Photonics, University of Erlangen-Nuernberg, Staudtstrasse 7/B2, 91058 Erlangen, Germany}
\email{*marco.ornigotti@uni-jena.de}

\begin{abstract}
We report on a new class of exact solutions of the scalar Helmholtz equation obtained by carefully engineering the form of the angular spectrum of a Bessel beam. We consider in particular the case in which the angular spectrum of such generalized beams has, in the paraxial zone, the same radial structure as Laguerre-Gaussian beams. We investigate the form of these new beams as well as their peculiar propagation properties.
\end{abstract}

\ocis{(260.2110) Electromagnetic Optics; (260.6042) Singluar Optics; (260.3160) Interference}

 ] 

\noindent 
The research field of optical beams, i.e., the study of electromagnetic field configurations that propagate mainly along a preferred direction and obey either the Helmholtz equation or its paraxial form, has been very flourishing since the invention and development of the Laser in the 1960s by Maiman et al. \cite{maiman, townes, russo}. Among the different properties and degrees of freedom of an optical beam that have been investigated, a fairly large amount of literature was dedicated to the study of their spatial structure, namely on how the intensity (and therefore the energy) can be distributed across the plane transverse with respect to the propagation direction and how this intensity distribution propagates. Perhaps, the most famous configurations are the so-called Hermite-Gaussian and Laguerre-Gaussian beams \cite{svelto}, which are solutions of the paraxial wave equation in Cartesian and Cylindrical coordinates respectively, and represent the transverse structure of the electromagnetic field emitted by most laser.  During the past decade, however, many other solutions  have been investigated both in the scalar and vector regimes such as Bessel beams \cite{bessel1}, generalized Hermite-Gaussian and Laguerre-Gaussian beams \cite{banders1,banders2,banders3}, Ince-Gaussian beams \cite{banders4}, Hypergeometric modes \cite{hyper1}, Airy beams \cite{airy1,airy2}, accelerating beams \cite{airy3}, Helmholtz-Gauss \cite{banders5} and Laplace-Gauss \cite{banders6} beams, just to name a few. Recently, new kind of solutions of the paraxial and nonparaxial equations were proposed in the form of propagation-invariant beams based on the diffraction pattern of cusp caustics (Pearcey function) \cite{dennis1}, beams with quantum pendulum spectra \cite{dennis2} and polynomial solutions in the spatial variables \cite{dennis3}. On the experimental side, unconventional field pattern in suitably modified cavities known as Throcloidal beams have also been reported\cite{troclo1,troclo2}. All these examples, however, have a common feature: they all represent solutions of the paraxial (or nonparaxial) equation in a determinate coordinate system or with suitable boundary conditions. This, however, is not the only possible direction to explore if one wants to build new classes of optical beams.

In this Letter, we present a new class of solutions of the Helmholtz equation obtained by means of a completely new and different approach: the engineering of the angular spectrum of a Bessel beam. In particular, we show how to construct generalized Bessel beams possessing a radial structure that is fully equivalent (in the paraxial region) to the one of Laguerre-Gaussian (LG) beams. To do this, we replace the Dirac delta function in the angular spectrum of a Bessel beam with a more general one which connects the zeros of the Delta function with the positions of the radial maxima of a LG beam. As a result, one obtains a superposition of Bessel beams with different angular apertures $\vartheta_0$, whose value are uniquely determined by the position of the radial maxima of the corresponding LG beam.

We begin our analysis by briefly recalling the definition of  a scalar monochromatic Bessel beam characterized by a cone angle aperture $\vartheta_0$, whose electric field can be written in the following form: 
\beq\label{eq1}
E(x,y,z)=J_l(K_0R)e^{il\theta}e^{iz\sqrt{k_0^2-K_0^2}},
\eeq
where $l$ is an integer number that fixes the value of the orbital angular momentum (OAM) of the beam, $R=\sqrt{x^2+y^2}$, $\tan\theta=y/x$, $K_0=k_0\sin\vartheta_0$ and $J_l(x)$ is the $l$-th order Bessel function of the first kind.  For the purposes of this Letter, it is convenient to calculate the angular spectrum associated to this field distribution as \cite{mandelWolf}:
\beq\label{eq2}
A(k_x,k_y)=\frac{1}{2\pi}\int E(x,y,0)e^{i\mathbf{K}\cdot\mathbf{R}}d^2K,
\eeq
where $\mathbf{K}=k_x\uvett{x}+k_y\uvett{y}$ and $d^2K=dk_xdk_y$. Upon substituting Eq. \eqref{eq1} into \eqref{eq2} we obtain the well-known angular spectrum representation of a Bessel beam, as
\beq\label{eq3}
A(k_x,k_y)=\frac{1}{i^lK_0}\delta(K-K_0)e^{il\phi},
\eeq
where $K=\sqrt{k_x^2+k_y^2}$ and $\tan\phi=k_y/k_x$. It is important to remark that the diffractionless nature of a Bessel beam is a direct consequence of its singular angular spectrum. Suppose now to operate the following replacement:
\beq\label{deltaF}
\frac{1}{K_0}\delta(K-K_0)\rightarrow\sum_{n=1}^NC_n\delta(K-K_n),
\eeq
where $C_n$ are arbitrary constants and the quantities  $K_n$ are in general arbitrary subjected to the only constrain that $K_n=k_0\sin\vartheta_n<k_0$. By inverting Eq. \eqref{eq2},  the electric field distribution associated with the  general angular spectrum \eqref{deltaF} assumes the form:
\beq\label{eq6}
E_l(x,y,z)=\sum_{n=1}^NC_nJ_l(K_nR)e^{iz\sqrt{k_0^2-K_n^2}}e^{il\theta}.
\eeq
\begin{figure}[!t]
\begin{center}
\includegraphics[width=0.5\textwidth]{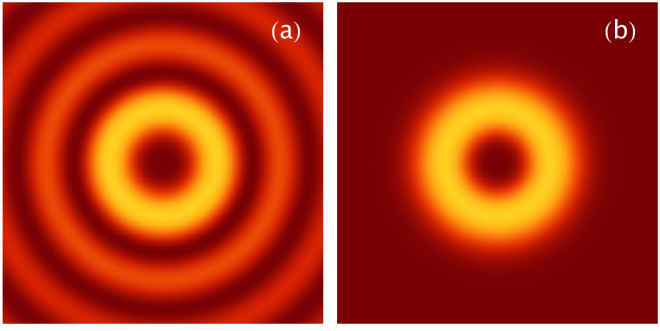}
\caption{Comparison between the transverse intensity distribution of (a) Bessel-like beam $E_2(x,y,0)$ and (b) the LG-beam $LG_0^2(x,y)$. Note how the central lobe of the Bessel -like beam has the same radial structure of the LG-beam. The axes of both graphs span the interval $[-3,3]$ in units of the beams waist $w_0$ of the LG-beam.}
\label{fig1}
\end{center}
\end{figure}
Equation \eqref{eq6} is still an exact solution of Helmholtz equation, and represents a whole new family of Bessel-like beams. Note that since $K_n=k_0 \sin\vartheta_n$, this solution corresponds in general to the superposition of $N$ different Bessel beams, each one characterized by its own angular aperture $\vartheta_n$. The form and properties of these new beams are determined by specifying the explicit form of $K_n$ in Eq. \eqref{deltaF}. In the present work, we construct a new class of exact solutions of the Helmholtz equation, which we call two-index Bessel beams, by requiring that the radial maxima in the paraxial region of these new beams are matched with the radial maxima of a LG-beam. This requirement translates then in a condition that determines the quantities $C_n$ and $K_n$ in Eq. \eqref{deltaF}. As it is well known \cite{svelto}, LG-beams are characterized by two indices: a radial index $p$ and an azimuthal index $l$, the latter quantifying the OAM carried by the beam. Bessel beams, on the other hand, only posses the azimuthal index $l$, as can be seen from Eq. \eqref{eq1}. However, the position of the radial zeros of a LG-beam depend on both the $p$ and $l$ index, and therefore applying as constrain that  the radial maxima of Eq. \eqref{eq6} correspond to the radial maxima of a LG-beam results in generating Bessel beams with 2 indices, one ($l$) related to the amount of OAM carried by the beam itself, and the other one ($p$) related to their radial structure. Notice that we implicitly assumed that both the LG- and Bessel-beams have the same azimuthal index $l$. Although this is not required by our constrain, without loss of generality we can limit our analysis to this case. To understand how to apply this constrain, let us first analyze the case $p=0$. The position of the radial maxima of a $l$-th order Bessel beams are found as the points in which its derivative vanishes, namely
\beq\label{besselroot}
J_{l-1}(\xi)-J_{l+1}(\xi)=0,
\eeq
where $\xi=K_0 R$. Since we intend to establish a direct connection between the structure of the two beams in the paraxial region, we only consider the smallest root of Eq. \eqref{besselroot} and we name it $\xi_l$.  The position of the radial maximum of a LG-beam with $p=0$ is instead given by $R=w_0\sqrt{|l|/2}$; if we define $\eta=\sqrt{2}R/w_0$, being $w_0$ the waist of the beam, we can introduce the dimensionless parameter $\eta_l=\sqrt{|l|}$. This result is immediate to verify since for $p=0$ we have $L_0^{|l|}(\eta)=1$ and the radial part of a LG-beam reduces to $\eta^{|l|}\exp{(-\eta^2/2)}$. The values of $K_n$ to put in Eq. \eqref{eq6} are then found by requiring that the position of the maxima given by $R_l=\xi_l/K_n$ for the generalized beam in Eq. \eqref{eq6} and $R_l=w_0\sqrt{|l|/2}$ of the LG-beam coincide. This constrain brings to the following result:
\beq\label{cond1}
\sin\vartheta_l=\frac{\xi_l}{\sqrt{2|l|}}\theta_0,
\eeq
where $\theta_0=2/(k_0w_0)$ is the beam divergence associated to the LG-beam \cite{svelto}. In this case, as we choose the azimuthal index of the two beams to be the same, we have a single index $l$ and the summation in Eq. \eqref{eq6} contains only one term. The result for $p=0$ is therefore just a single Bessel beam whose angular aperture is determined as a function of the amount of OAM carried by the beam itself, namely
\beq
E_l(x,y,z)=C_lJ_l(k_0\sin\vartheta_lR)e^{ik_0z\cos\vartheta_l},
\eeq
where we have choses the arbitrary constant $C_l=\eta_l^{|l|}\exp{(-\eta_l^2/2)}$ to be the value of the LG-beam at its radial maximum. The reason of this choice will be clear later, when we will discuss the general case with $p\neq 0$. A comparison of this solution with the correspondent LG-beam is shown in Fig. \ref{fig1}. As it can be seen, this operation for $p = 0$ only results in a rescaling (upon the beam waist $w_0$ of the correspondent LG-beam) of a
normal Bessel beam.  Before turning our attention to the case $p\neq 0$, it is instructive to analyze the case $l=0=p$. In this case, in fact, since the smallest root of Eq. \eqref{besselroot} is $\xi_0=0$, the condition expressed by Eq. \eqref{cond1} has no sense anymore. In this particular case, however, we can still rescale the $0$-th order Bessel beam with respect to the LG-beam waist $w_0$ by simply requiring that $\sin\vartheta_0=\theta_0$.

We now turn our attention to the most interesting case of $p\neq 0$. In general, a LG-beam with radial index $p$ possesses $p+1$ radial maxima \cite{nist}. In this case, therefore, we will obtain $N=p+1$ different expressions of $K_n$, whose value depend upon the choice of both indices $l$ and $p$. Each term in the summation \eqref{eq6} therefore represents  a Bessel beam whose first radial maximum coincides with the $n$-th radial maximum of the correspondent LG-beam. The consequence of this is that the beam described by Eq. \eqref{eq6} depends upon two indices: the index $l$ that defines its OAM (as it is done for a standard Bessel beam), and the index $p$ that determines the radial structure of the beam (in the same way that happens for a LG-beam).  We call this new solution of the Helmholtz equation a two-index Bessel beam. 

\begin{figure}[!t]
\begin{center}
\includegraphics[width=0.5\textwidth]{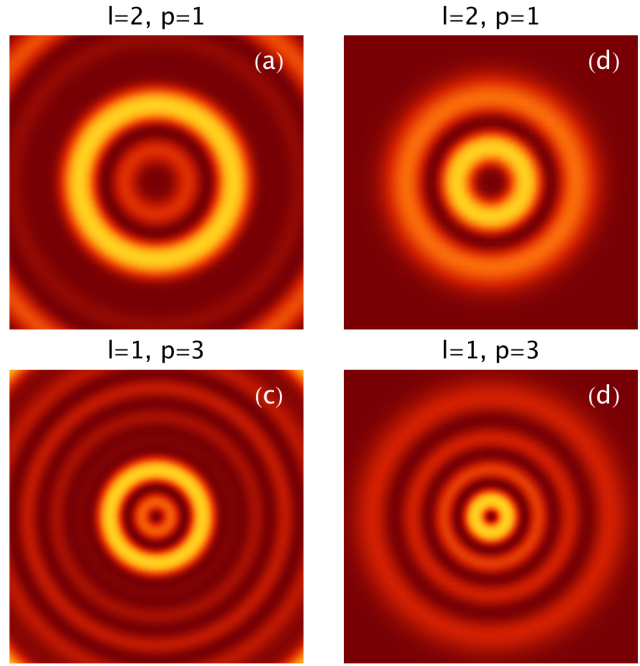}
\caption{Comparison between the transverse intensity distribution of a two-index Bessel beam $E_p^l(x,y,0)$ (left column) and the correspondent LG-beam $LG_p^l(x,y,0)$ (right column) in $z=0$ for. The values of the index $l$ and $p$ are reported n top of each figure. As can be noted, the number of rings of the two-index Bessel beams in the paraxial regime equals the number of rings of the correspondent LG-beams. The axes of all the graphs span the interval $[-3,3]$ in units of the beams waist $w_0$ of the LG-beam.}
\label{fig2}
\end{center}
\end{figure}

To find the values of $K_n$, we recall that the position of the maxima of a LG-beam with $p\neq 0$ are found by finding the roots of the following equation:
\beq\label{laguerreroot}
|l|L_p^{|l|}(\eta^2)-\eta^2\left[2L_{p-1}^{|l|+1}(\eta^2)+L_p^{|l|}(\eta^2)\right]=0,
\eeq
where as before $\eta=\sqrt{2}R/w_0$. The condition given by Eq. \eqref{cond1} now generalizes to
\beq
\sin\vartheta_n=\frac{\xi_l}{\sqrt{2}\eta_n}\theta_0,
\eeq
where we indicated with $\eta_n$ the $n$-th root of Eq. \eqref{laguerreroot}. Inserting this result into Eq. \eqref{eq6} brings to
\beq\label{Elp}
E_{pl}(x,y,z)=\sum_{n=1}^{p+1}C_nJ_l(k_0\sin\vartheta_nR)e^{ik_0z\cos\vartheta_n}e^{il\theta}.
\eeq
It is useful at this point to discuss the form of the weight coefficients $C_n$. In order to make the correspondence complete, in fact, it is not sufficient to require that the first maximum of each Bessel beam in the summation above corresponds to the $n$-th maximum of the correspondent LG-beam, but it is also necessary that the relative weight between the maxima is also reproduced. If this condition is fulfilled, in fact, each term of the summation will contribute with the proper weight to the construction of the generalized beam. We therefore require that the coefficients $C_n$ are given by $C_n=\eta_n^{|l|}\exp^{(-\eta_n/2)}L_p^{|l|}(\eta_n^2)$. The transverse intensity distribution of $E_p^l(x,y,0)$ for some values of the two indices $p$ and $l$, together with their comparison with the correspondent LG-beams are depicted in Fig. \ref{fig2}. This is the main result of our Letter. Two-index Bessel beams constitute a novel class of exact solutions of the Helmholtz equation, and they can be represented as a superposition of conventional Bessel beam whose cone angle $\vartheta_n$ is chosen (in this particular case) in such a way to give to the two-index Bessel beam the same paraxial structure of a LG-beam. 
We conclude our analysis of this new class of beams by investigating their propagation features. Since the two-index Bessel beams are written in Eq. \eqref{eq6} as a superposition of conventional Bessel beams, it is interesting to study the evolution of the intensity of such beams along the $z$-direction, as interference between the various components of the beam can give rise to interesting features and patterns. If we define $k_{zn}=k_0\cos\vartheta_n$ as the propagation constant of each Bessel beam component,  $\kappa_n=w_0K_n$ and we introduce the dimensionless radial coordinate $\rho=R/w_0$,  the total intensity of a two-index Bessel beam can be written as
\barr\label{intensity}
I_{pl}(x,y,z)=|E_{pl}(x,y,z)|^2=\sum_{j=1}^{p+1}C_j^2J_l^2\left(\kappa_n\rho\right)\nonumber\\
+2\sum_{n=1}^{p+1}\sum_{j= n+1}^{p+1}\Lambda_{j,n}\cos\left[(k_{zj}-k_{zn})z\right],
\earr
where $\Lambda_{j,n}=C_jC_nJ_l\left(\kappa_n\rho\right)J_l\left(\kappa_j\rho\right)$. Let us analyze three distinct cases. If $p=0$, the second summation in Eq. \eqref{intensity} is zero and Eq. \eqref{intensity} reduces to
\beq
I_{0l}(x,y,z)=C_l^2J_l^2\left(\kappa_n\rho\right),
\eeq
and the corresponding beams are diffraction-less. As it was pointed out before, the case $p=0$ reduces the two-index Bessel beam to the conventional Bessel beam with rescaled coordinates with respect to the scaling parameter $w_0$. If now we put $l=0$ and $p$ arbitrary, we have that $\kappa_n=2=\kappa_j$ and $k_{z,j}-k_{z,n}=0$, therefore the intensity is given by
\beq
I_{p0}(x,y,z)=\Gamma J_0^2(2\rho),
\eeq
where $\Gamma=\sum_{j=1}^{p+1}C_j^2+2\sum_{n=1}^{p+1}\sum_{j=n+1}^{p+1}C_nC_j$. 

\begin{figure}[!t]
\begin{center}
\includegraphics[width=0.5\textwidth]{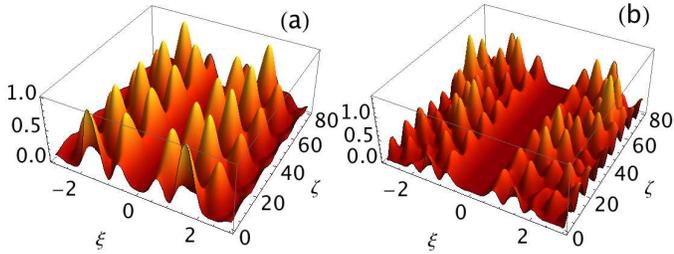}
\caption{Section (at $y=0$) of the evolution of the normalized intensity  of the two-indices Bessel beam $E_{pl}(\xi,0,\zeta)$ as a function the normalized coordinate $\xi=x/w_0$ and propagation direction $\zeta=z/z_R$ (being $z_R=k_0w_0^2/2$ the Rayleigh range of the beam), for (a) $(p,l)=(1,2)$  and (b) $(p,l)=(3,4)$. The beams are propagating along $\zeta$ with a periodic pattern, whose period depends essentially on $p$. As can be seen from the figures, in fact, as $p$ increases, the periodicity of  the beam along the propagation direction increases accordingly.}
\label{fig3}
\end{center}
\end{figure}

Two-index Bessel beams with $l=0$ are also diffraction-less. The last case to consider is the general case where $l\neq 0$ and $p\neq 0$. In this case, the intensity distribution is periodic with respect to $z$, with a period that is essentially determined by the index $p$ through the term $k_{zj}-k_{zn}$ inside the cosine function and through the fact that $p$ defines also the number of such cosine terms that have to be summed together. However, although formally this solution is not diffraction-less anymore in the traditional sense (i.e. the intensity distribution is not anymore $z$-independent), Fig. \ref{fig3} shows that during propagation, the intensity distribution given by Eq. \eqref{intensity} reproduces itself periodically, thus exhibiting a quasi-diffraction-free behavior.

In conclusion, we have introduced a new class of solutions of the scalar Helmholtz equation based on a generalization of the angular spectrum of a Bessel beam according to Eq. \eqref{deltaF}. We have show that by choosing $K_n$ in such a way that the maxima  of Eq. \eqref{eq6} correspond to the maxima of a LG-beam it is possible to obtain a new class of solutions having two indices that we named two-index Bessel beams. We have shown that these beams are parametrized upon the parameter $w_0$ that in this case is the beam waist of the corresponding LG-beam and that their propagation properties depend upon the choice of the index couple $(p,l)$, and that this class of beams contains full diffraction-less beams as well as periodically self-reproducing beams.


\pagebreak

\section*{Informational Fifth Page}

\end{document}